\documentclass[12pt]{article}
\usepackage{graphicx}
\usepackage{amssymb}
\usepackage{amsmath}

\begin{document}

\begin{titlepage}

\begin{flushright}
ICRR-Report-602-2011-19\\
IPMU~11-0213
\end{flushright}

\vskip 1.35cm

\begin{center}

{\large 
{\bf Hubble Induced Mass in Radiation Dominated Universe} 
}

\vskip 1.2cm

Masahiro Kawasaki$^{a,b}$
and
Tomohiro Takesako$^a$ \\

\vskip 0.4cm

{ \it$^a$Institute for Cosmic Ray Research,
University of Tokyo, Kashiwa 277-8582, Japan}\\
{\it $^b$Institute for the Physics and Mathematics of the Universe,
University of Tokyo, Kashiwa 277-8568, Japan}\\

\date{\today}

\begin{abstract} 
We reconsider the effective mass of a scalar field 
which interact with visible sector via Planck-suppressed coupling in supergravity framework.
We focus on the radiation-dominated (RD) era after inflation.
In this era, the effective mass is given by thermal average of interaction terms.
To make our analysis clear, we rely on Kadanoff-Baym equations to evaluate the thermal average.
We find that, in RD era, a scalar field acquires the effective mass of the order of $H$. 
\end{abstract}

\end{center}
\end{titlepage}

\section{Introduction}
\label{sec:intro}
Supergravity provides many interesting phenomena in the early universe where Planck-suppressed operators play an important role.
In inflationary era, a scalar field generally obtain Hubble-induced effective mass of order $H$ through supergravity effect~\cite{hep-ph/9405389}.
In a word, this Hubble-induced mass is originated from the energy of inflation which breaks supersymmetry~\cite{Dine:1995uk}.
After inflationary era, reheating process may occur and radiation-dominated (RD) era follows.
Then a question arise: is there any source for the effective mass of the order of $H$ after inflation? 
There is a possibility that the thermal plasma in RD era provides the source as expected in Ref.~\cite{Dine:1995uk}, 
since the inflation energy seems to have converted to the energy of the plasma through reheating process.
Our main purpose of this study is to answer the above question.

We consider two complex scalar fields $\phi$ and $\chi$, whose masses are originally ({\it i.e.}, at zero temperature) much smaller than the Hubble scale $H$, in supergravity framework.
Here, $\phi$ is assumed to be decoupled from the thermal bath, whereas $\chi$ is in equilibrium with the bath in RD era.
It is assumed that these two fields $\phi, \chi$ interact with each other via the non-minimal K$\ddot{\rm{a}}$hler potential given by
\begin{equation}\label{eq:kahler}
\begin{split}
K = |\phi|^2 + |\chi|^2 + c~\frac{|\phi|^2  |\chi|^2}{M_{\text{P}}^2},
\end{split}
\end{equation}
where $M_{\text{P}} \simeq 2.4 \times 10^{18}~\mathrm{GeV}$ is the reduced Planck mass and $c = \mathcal{O} (1)$ is a model-dependent parmameter. 
Then, the kinetic term of $\chi$ has the form~\cite{350988}:
\begin{equation}\label{eq:kin}
\begin{split}
\mathcal{L}^{\chi}_{\text{kin.}}
&= K_{\chi \bar \chi} \partial_{\mu} \chi^* \partial^{\mu} \chi \\
&= \left( 1 + c~\frac{|\phi|^2}{M_{\text{P}}^2} \right) \partial_{\mu} \chi^* \partial^{\mu} \chi.
\end{split}
\end{equation}

In the following, we consider the effecive mass-squared of the scalar field $\phi$, $\tilde m_{\phi}^2$, especially in RD era after the inflationary era.
In this era, when the Hubble-induced mass due to the inflation potential disappear, we are interested in what a value the effective mass-squared $\tilde m_{\phi}^2$ takes.
In Ref.~\cite{hep-ph/0402174}, it is insisted that the effective mass-squared takes a value much smaller than the Hubble scale:
$\tilde m_{\phi}^2 \sim \frac{m_{\chi}^2}{T^2} H^2 \ll H^2$, where $m_{\chi}$ is the zero temperature mass of $\chi$ and $T$ is the temperature of the thermal bath.

However, the following argument seems possible~\cite{Asaka:1999yc}.
From the kinetic term Eq.~(\ref{eq:kin}), 
the effective mass-squared $\tilde m_{\phi}^2$ originated from the $\phi\,$-$\,\chi$ Planck-suppressed interaction is generally written as
\begin{equation}
\begin{split}
\tilde m_{\phi}^2 |_{\text{kin.}} = - \frac{c}{M_{\text{P}}^2} \langle  \partial_{\mu} \chi^* \partial^{\mu} \chi \rangle.
\end{split}
\end{equation}
Hereafter, $\langle \cdots \rangle \equiv \mathrm{tr} (\hat \rho \cdots) / \mathrm{tr} (\hat \rho)$ represents an expectation value with the density matrix $\hat \rho$. 
For a thermal equilibrium system, $\langle \cdots \rangle$ gives the thermal average with $\hat \rho = \mathrm{exp}(- \beta \hat{\mathcal{H}} )$,
where $\beta = 1/T$ is the inverse temperature and $\hat{\mathcal{H}}$ is the Hamiltonian of the system.
Therefore, the effective mass-squared of $\phi$ originated from Eq.~(\ref{eq:kin}) is determined by the thermal average $\langle  \partial_{\mu} \chi^* \partial^{\mu} \chi \rangle$ in RD era.
Using equation of motion for $\chi$, we naively estimate the effective mass-squared as\footnote{
Such an estimate seems to have been raised in Ref.~\cite{Dine:1995uk}.
}
\begin{equation}\label{eq:naive}
\begin{split}
\tilde m_{\phi}^2 |_{\text{kin.}}
&\simeq \frac{c}{M_{\text{P}}^2} \langle  \chi^* \square \chi \rangle \\
&\simeq - \frac{c m_{\text{th}}^2}{M_{\text{P}}^2} \langle  \chi^* \chi \rangle \\
&\simeq - c g^2 H^2,
\end{split}
\end{equation}
where $g$ is a coupling strength of the scalar field $\chi$ to the thermal bath.
Here, thermal mass $m_{\text{th}} \simeq g T$ for $\chi$, $\langle  \chi^* \chi \rangle \simeq T^2$, and $T^4 \simeq H^2 M^2_{\text{p}}$ are used. 
In Eq.~(\ref{eq:naive}), the nontrivial equalities are  the first and second line, namely, 
it is ambiguous whether or not we can use equation of motion and the thermal mass in the equalities.

There is another naive estimation.
Let us directly evaluate $\langle  \partial_{\mu} \chi^* \partial^{\mu} \chi \rangle$ by using the expansion of the scalar field $\chi$:
\begin{equation}
\begin{split}
\chi (x) = \int \frac{\mathrm{d}^3 \bold{k}}{(2 \pi)^3 2 \omega_{\bold{k}}} \left( a_{\bold{k}} \mathrm{e}^{- i k \cdot x} + a_{\bold{k}}^{\dagger} \mathrm{e}^{ i k \cdot x} \right)
\end{split}
\end{equation}
where $\omega_{\bold{k}} = \sqrt{k^2 + M_{\chi}^2}$, $M_{\chi}$ is a kinetic mass of $\chi$, and $a_{\bold{k}} (a_{\bold{k}}^{\dagger})$ is the annihilation (creation) operator.
Then, 
\begin{equation}\label{eq:naive2}
\begin{split}
\tilde m_{\phi}^2 |_{\text{kin.}}
&=  - \frac{c}{M_{\text{P}}^2} \langle  \partial_{\mu} \chi^* \partial^{\mu} \chi \rangle \\
&=  - \frac{c M_{\chi}^2}{M_{\text{P}}^2} \int \frac{\mathrm{d}^3 \bold{k}}{(2 \pi)^3 2 \omega_{\bold{k}}} \frac{1}{\mathrm{e}^{\beta \omega_{\bold{k}} } - 1},
\end{split}
\end{equation}
where we have used $k^2 = M_{\chi}^2$.
One may consider $M_{\chi}$ is the zero temperature mass $M_{\chi} = m_{\chi} \ll T$, leading to $\tilde m_{\phi}^2 \ll H^2$.
On the other hand, as the scalar field $\chi$ is in the thermal bath, another may insist $\chi$ acquire a thermal mass and $M_{\chi} = m_{\text{th}} \simeq g T$,
leading to $\tilde m_{\phi}^2 \simeq H^2$.
These considerations should be confirmed by using a reliable formulation.

Therefore, the main purpose of the present paper is to answer the question
whether $\langle  \partial_{\mu} \chi^* \partial^{\mu} \chi \rangle \simeq g^2 T^4$ is correct or not.
The strategy is to express $\langle  \partial_{\mu} \chi^* \partial^{\mu} \chi \rangle$ by a solution of Kadanoff-Baym (KB) equations~\cite{KB}
which naturally describe thermal effects in Green functions.
By using a solution of KB equations, we could answer the above question quantitatively. 

The construction of this paper is as following: 
in section~\ref{sec:KBeq}, we briefly describe the KB equations and their solution.
In section~\ref{sec:eff1}, we express $\langle  \partial_{\mu} \chi^* \partial^{\mu} \chi \rangle$ by a solution of KB equations and 
derive a formula for the effective mass-squared of the scalar field $\phi$ with the quasiparticle approximation for the scalar field $\chi$.
Section~\ref{sec:conc} is devoted to conclusion.

\section{Kadanoff-Baym equations for a real scalar field}
\label{sec:KBeq}

In this section, we briefly review the formalism of KB equations~\cite{hep-ph/0409233, Anisimov:2008dz, Garny:2009ni}, 
which we use to express $\langle  \partial_{\mu} \chi^* \partial^{\mu} \chi \rangle$ in the next section.
Although KB equations are usually applied to non-equilibrium system, we can use them also to equilibrium system~\cite{Anisimov:2008dz, Garny:2009ni}.
In prospect of extension our analysis to non-equilibrium situations in future, we use KB equations to an equilibrium setup in the next section. 

We decompose the complex scalar field $\chi$ as 
\begin{equation}
\begin{split}
\chi = \frac{1}{\sqrt{2}} \left( \chi_1 + i \chi_2 \right), 
\end{split}
\end{equation}
where $\chi_1$ and $\chi_2$ are real scalar fields.
In the following, we treat only the real scalar field $\chi_1$ for simplicity.
The generalization to the complex scalar field $\chi$ is straightforward.

In Minkowski space-time, the time-ordered Green function for an interacting real scalar field $\chi_1$ in the Keldysh formalism~\cite{3567}, $\Delta_C (x_1, x_2)$, satisfies the following Dyson-Schwinger equation:
\begin{equation}\label{eq:Shwinger-Dyson}
\begin{split}
\left( \square_1 + m^2_{\chi} \right) \Delta_C (x_1, x_2) + \int_C \mathrm{d}^4 x' ~i \Pi_C (x_1, x')  \Delta_C (x', x_2) = - i \delta^{(4)}_C (x_1 - x_2),
\end{split}
\end{equation}
where $C$ denotes the Keldysh contour and $\Pi_C (x_1, x_2)$ is the self-energy of $\chi_1$.
Here, $\Delta_C (x_1, x_2) = \theta_C (t_1 - t_2) \Delta^> (x_1, x_2) + \theta_C (t_2 - t_1) \Delta^< (x_1, x_2)$,
$\Delta^> (x_1, x_2) = \langle \chi (x_1) \chi (x_2) \rangle$ and $\Delta^< (x_1, x_2) =  \langle \chi (x_2) \chi (x_1) \rangle$.
$\delta_C (t)$ and $\theta_C (t)$ are the delta function and theta function on the contour $C$, respectively.
Now, we define
\begin{equation}
\begin{split}
&\Delta^- (x_1, x_2) = i \left( \Delta^> (x_1, x_2) - \Delta^< (x_1, x_2) \right) = i \langle [ \chi (x_1) , \chi (x_2) ] \rangle, \\
&\Delta^+ (x_1, x_2) = \frac{1}{2} \left( \Delta^> (x_1, x_2) + \Delta^< (x_1, x_2) \right) = \frac{1}{2} \langle \{ \chi (x_1) , \chi (x_2) \} \rangle,
\end{split}
\end{equation}
which are called spectral function and statistical propagator, respectively.
For the self-energy of $\chi_1$, we decompose $\Pi_C (x_1, x_2)$ as $\Pi_C (x_1, x_2) = \theta_C (t_1 - t_2) \Pi^> (x_1, x_2) + \theta_C (t_2 - t_1) \Pi^< (x_1, x_2)$
and we define
\begin{equation}
\begin{split}
&\Pi^- (x_1, x_2) = i \left( \Pi^> (x_1, x_2) - \Pi^< (x_1, x_2) \right) , \\
&\Pi^+ (x_1, x_2) = \frac{1}{2} \left( \Pi^> (x_1, x_2) + \Pi^< (x_1, x_2) \right).
\end{split}
\end{equation}

Assuming spatial homogeneity, the Dyson-Schwinger equation~(\ref{eq:Shwinger-Dyson}) is reduced to the following 
coupled equations for the spatial Fourier transforms $\Delta^-_{\bold{p}}$ and $\Delta^+_{\bold{p}}$\footnote{
Strictly speaking, Eq.~(\ref{eq:12KB}) is derived assuming a Gaussian initial condition for the Green functions.
When a non-Gaussian initial condition is taken into account, there are additional terms to the 2nd KB equation~\cite{Garny:2009ni}.
}
:
\begin{equation}\label{eq:12KB}
\begin{split}
&\left( \partial_{t_1}^2 + \omega_{\bold{p}}^2 \right) \Delta^-_{\bold{p}} (t_1, t_2) + \int_{t_2}^{t_1} \mathrm{d} t' ~\Pi^-_{\bold{p}} (t_1, t') \Delta^-_{\bold{p}} (t', t_2) = 0, \\
&\left( \partial_{t_1}^2 + \omega_{\bold{p}}^2 \right) \Delta^+_{\bold{p}} (t_1, t_2) + \int_{t_i}^{t_1} \mathrm{d} t' ~\Pi^-_{\bold{p}} (t_1, t') \Delta^+_{\bold{p}} (t', t_2) =  \int_{t_i}^{t_2} \mathrm{d} t'  ~\Pi^+_{\bold{p}} (t_1, t') \Delta^-_{\bold{p}} (t', t_2),
\end{split}
\end{equation}
where $\omega^2_{\bold{p}} = \bold{p}^2 + m^2_{\chi}$, and $t_i$ is an initial time.
These equations are called 1st and 2nd Kadanoff-Baym equations, respectively.

In the following, we also assume time translational invariance as well as the spatial homogeneity for simplicity.
These assumptions are justified when a thermal equilibrium system is concerned\footnote{
For non-equilibrium systems, see Refs.~\cite{hep-ph/0409233, Anisimov:2008dz, Drewes:2010pf, arXiv:1111.4594} for example.
}
. 
In this case, the spatial Fourier transform $\Delta^-_{\bold{p}} (t_1 , t_2)$ depends only on the time difference $y = t_1 - t_2$. 
Moreover, from the equal-time commutation relation of $\chi_1$, the boundary condition for $\Delta_{\bold{p}}$ is given by
\begin{equation}\label{eq:p-boundary}
\begin{split}
&\Delta^-_{\bold{p}} (y) |_{y=0} = \partial_y^2 \Delta^-_{\bold{p}} (y) |_{y=0} = 0, \\
&\partial_{y } \Delta^-_{\bold{p}} (y) |_{y=0} = 1.
\end{split}
\end{equation}
Then, the solution of the 1st KB equation is given by~\cite{Anisimov:2008dz}
\begin{equation}\label{eq:spec}
\begin{split}
&\Delta^-_{\bold{p}} (y) = i \int_{- \infty}^{\infty} \frac{\mathrm{d} \omega}{2 \pi} \mathrm{e}^{- i \omega y} \rho_{\bold{p}} (\omega), \\
& \rho_{\bold{p}} (\omega) = \frac{- 2~ \mathrm{Im}~ \Pi^R_{\bold{p}} (\omega) + 2 \omega \epsilon}{\left[ \omega^2 - \omega_{\bold{p}}^2 - \mathrm{Re}~\Pi^R_{\bold{p}} (\omega) \right]^2 + \left[ \mathrm{Im}~\Pi^R_{\bold{p}} (\omega) - \omega \epsilon \right]^2} ,
\end{split}
\end{equation}
where $\Pi^R_{\bold{p}} (\omega)$ is the Fourier transform of $\Pi^R (x_1, x_2) = \theta (t_1 - t_2) \Pi^- (x_1, x_2)$ and $\epsilon \to + 0$.
From Eq.~(\ref{eq:spec}), 
we can see that the spectral function of $\chi_1$ has thermally corrected poles in RD era, 
namely, the dispersion relations of the poles differ from the ones in zero temperature.

\section{Hubble-induced mass from kinetic term}
\label{sec:eff1}

In this section, we derive a formula for the effective mass of the scalar field $\phi$.
We assume here spatial homogeneity and isotropy of the background metric.
We also take $m_{\chi} = 0$ for simplicity, although the following argument can be applied for nonzero $m_{\chi}$ with $m_{\chi} \ll H$.

\subsection{Formulation with quasiparticle approximation}
Using the KB equations described in the previous section,
let us evaluate the expectation value $\langle  \partial_{\mu} \chi_1 \partial^{\mu} \chi_1 \rangle$ when $\chi_1$ is in equilibrium with the thermal bath.
Here, $\chi_1$ is assumed to be in non-equilibrium with the bath at an initial time $t_i$ and then thermalized at a late time. 
Such a situation is discussed with the usage of Eq.~(\ref{eq:12KB}) in literatures (see \cite{hep-ph/0409233, Anisimov:2008dz} for example).
We evaluate $\langle  \partial_{\mu} \chi_1 \partial^{\mu} \chi_1 \rangle$ at a late time so that $\chi_1$ under consideration is in thermal equilibrium.
Then, Eq.~(\ref{eq:12KB}) can be applied to the evaluation of $\langle  \partial_{\mu} \chi_1 \partial^{\mu} \chi_1 \rangle$, while $\chi_1$ is in thermal equilibrium.
The thermalization of $\chi_1$ is assumed to take much less time than the Hubble expansion time scale,
and the effect of Hubble expansion rate is effectively included in the plasma temperature $T$\footnote{
KB equations in curved space-time has been studied in Ref.~\cite{arXiv:0807.4551}.
}
.
Moreover, assuming the thermal bath is large enough, we neglect the effect of  $\phi\,$-$\chi_1$ interaction Eq.~(\ref{eq:kahler}) to the bath.

First of all, we express this expectation value by the statistical propagator for the real scalar field $\chi_1$.  
For this purpose, we note the following equation:
\begin{equation}
\begin{split}
\langle  \partial_{\mu} \chi_1 (x) \partial^{\mu} \chi_1 (x) \rangle
&=  \partial_{\mu}^{x_1} \partial^{x_2 \mu} \Delta^+ (x_1, x_2) |_{x_1=x_2=x}.
\end{split}
\end{equation}
Since $\chi_1$ is in thermal equilibrium, 
the spectral function and the statistical propagator for $\chi_1$ depend only on the difference of two points: $\Delta^{\pm} (x_1, x_2) = \Delta^{\pm (\text{eq})} (x_1- x_2)$.
We firstly use the spatial Fourier transform as
\begin{equation}\label{eq:kin-2}
\begin{split}
\langle  \partial_{\mu} \chi_1 (x) \partial^{\mu} \chi_1 (x) \rangle
&=  \partial_{\mu}^{x_1} \partial^{x_2 \mu} \int \frac{\mathrm{d}^3 \bold{p}}{(2 \pi)^3} \mathrm{e}^{+ i \bold{p} \cdot (\bold{x_1 -x_2})} \Delta^{+ (\text{eq})}_{\bold{p}} (t_1 - t_2) {\bigg |}_{x_1=x_2=x} \\
&=  \int \frac{\mathrm{d}^3 \bold{p}}{(2 \pi)^3} (- \partial_y^2 - \bold{p}^2) \Delta^{+ (\text{eq})}_{\bold{p}} (y) {\bigg |}_{y=0},
\end{split}
\end{equation}
where $y=t_1-t_2$.

Next, we have to know the expression for $\Delta^{+ (\text{eq})}_{\bold{p}} (y)$.
Since the real scalar field $\chi_1$ is in thermal equilibrium, we can use the KMS relation~\cite{8651}:
\begin{equation}
\begin{split}
\Delta^{+ (\text{eq})}_{\bold{p}} (\omega)
&= \frac{- i}{2} \coth \left( \frac{\beta \omega}{2} \right) \Delta^{- (\text{eq})}_{\bold{p}} (\omega).
\end{split}
\end{equation}
In fact, in Ref.~\cite{Anisimov:2008dz}, it is discussed that the KMS relation is realized for a real scalar field like $\chi_1$ at a late time using Eq.~(\ref{eq:12KB}).
Then, we obtain the following expression:
\begin{equation}\label{eq:delta+}
\begin{split}
\Delta^{+ (\text{eq})}_{\bold{p}} (y) 
&= \int_{- \infty}^{\infty} \frac{\mathrm{d} \omega}{2 \pi} \mathrm{e}^{- i \omega y} \Delta^{+ (\text{eq})}_{\bold{p}} (\omega) \\
&= \int_{- \infty}^{\infty} \frac{\mathrm{d} \omega}{4 \pi} \mathrm{e}^{- i \omega y} \coth \left( \frac{\beta \omega}{2} \right) \rho_{\bold{p}} (\omega),
\end{split}
\end{equation}
where the relation $\Delta^{- (\text{eq})}_{\bold{p}} (\omega)= i \rho_{\bold{p}} (\omega)$ has been used.

Now, the problem is reduced to what a form the spectral function $\rho_{\bold{p}}$ takes, 
whose general form is already known from Eq.~(\ref{eq:spec}).
We take Eq.~(\ref{eq:spec}) as the basis of our study below.
Let us apply the quasiparticle approximation to the real scalar field $\chi_1$ in the thermal bath.
In this approximation, the interactions are assumed to be included in the thermally corrected effective masses of quasiparticles~\cite{Peshier:1995ty}.
Then, quasiparticles interact only weakly, and the imaginary parts of poles of the spectral function are assumed to be much smaller than the real counterparts.
Therefore, in the quasiparticle approximation of $\chi_1$, the spectral function Eq.~(\ref{eq:spec}) has the Breit-Wigner form:
\begin{equation}\label{eq:spec-app}
\begin{split}
\rho_{\bold{p}} (\omega) 
&\simeq \sum_{r=\pm} \frac{1}{\Omega_{\bold{p}}} \frac{r \Gamma_{\bold{p}} / 2}{(r \omega - \Omega_{\bold{p}})^2 + (\Gamma_{\bold{p}} / 2)^2}.
\end{split}
\end{equation}
There are four poles in the spectral function~(\ref{eq:spec-app}).
If we denote one of the poles as $\hat \Omega_{\bold{p}} = \Omega_{\bold{p}} - i \Gamma_{\bold{p}} / 2$, 
the spectral function $\rho_{\bold{p}} (\omega)$ has two poles $\hat \Omega_{\bold{p}}, - \hat \Omega_{\bold{p}}^*$ in the lower half complex-$\omega$ plane, 
and two poles $\hat \Omega_{\bold{p}}^*, - \hat \Omega_{\bold{p}}$ in the upper half plane.
Here, $\Omega_{\bold{p}}$ is the quasiparticle energy
and $\Gamma_{\bold{p}} = - \mathrm{Im}~\Pi_{\bold{p}}^R (\Omega_{\bold{p}}) / \Omega_{\bold{p}}$ is the quasiparticle width.
Then, it is easy to obtain $\Delta^{+ (\text{eq})}_{\bold{p}} (y)$ from Eq.~(\ref{eq:delta+}) by using the complex integration as\footnote{
There are other poles on imaginary axis of complex-$\omega$ plane arising from the factor $\coth(\beta \omega/2)$ in Eq.~(\ref{eq:delta+}).
These poles, however, converge on the origin of the plane in zero temperature limit, and are irrelevant to 1-particle state poles in zero temperature. 
Thus, in the following argument, we neglect these "thermal poles".
}
\begin{equation}\label{eq:delta+2}
\begin{split}
\Delta^{+ (\text{eq})}_{\bold{p}} (y) 
&= \mathrm{Re} \left\{ \frac{\mathrm{e}^{- i \hat \Omega_{\bold{p}} y} }{\Omega_{\bold{p}}} \left( \frac{1}{2} + n_B (\hat \Omega_{\bold{p}}) \right) \right\},
\end{split}
\end{equation}
where $n_B (\omega) = 1/ (\mathrm{e}^{\beta \omega} - 1)$ is the Bose-Einstein distribution function.

Now, we are in position to evaluate the expectation value $\langle  \partial_{\mu} \chi_1 (x) \partial^{\mu} \chi_1 (x) \rangle$.
Substituting Eq.~(\ref{eq:delta+2}) into Eq.~(\ref{eq:kin-2}), we obtain the following expression:
\begin{equation}\label{eq:kin-isotropic}
\begin{split}
\langle  \partial_{\mu} \chi_1 (x) \partial^{\mu} \chi_1 (x) \rangle
&=  \frac{1}{2 \pi^2}\int_0^{\infty} \mathrm{d} p~p^2~\mathrm{Re} \left\{ \frac{\hat \Omega_{p}^2 - p^2}{\Omega_p} \left( \frac{1}{2} + n_B (\hat \Omega_p) \right) \right\},
\end{split}
\end{equation}
where we have used spatial isotropy since $\chi_1$ is in thermal equilibrium.
We note that the time derivative in Eq.~(\ref{eq:kin-2}) picks up the thermally corrected poles, which
make sure the validity of substituting thermal mass in our naive estimate in Eqs.~(\ref{eq:naive}) and (\ref{eq:naive2}).

\subsection{Expectation value of kinetic term and the Hubble-induced mass of scalar field $\phi$}
In this subsection,
we write down more explicit expression for $\langle  \partial_{\mu} \chi_1 (x) \partial^{\mu} \chi_1 (x) \rangle$ 
in somewhat special but interesting case, so as to examine the estimate $\langle  \partial_{\mu} \chi^* (x) \partial^{\mu} \chi (x) \rangle \sim g^2 T^2$ quantitatively.

In the quasiparticle approximation for the real scalar field $\chi_1$ which is in thermal equilibrium, 
the width $\Gamma_p$ is assumed to be much smaller than the quasiparticle energy  $\Omega_p$.
In general, the dispersion relation for $\Omega_p$ has a complicated dependence on the momentum $\bold{p}$ 
and different from the one in vacuum~\cite{Bellac}.
However, for simplicity, we consider the case where $\hat \Omega_p$ has the following form:
\begin{equation}\label{eq:dispersion}
\begin{split}
\hat \Omega_p = \sqrt{p^2 + m_{\text{th}}^2} - i \Gamma_p / 2,~~\left( m_{\text{th}} \gg \Gamma_p \right).
\end{split}
\end{equation}
Such a form can be realized at the leading order if the self-energy of the real scalar field $\chi_1$ is dominated by, for example, the quartic interactions~\cite{Drewes:2010pf}.
In this case, the factor in Eq.~(\ref{eq:kin-isotropic}) becomes
\begin{equation}\label{eq:omega}
\begin{split}
\frac{\hat \Omega_{p}^2 - p^2}{\Omega_p}
&= \frac{m_{\text{th}}^2}{\sqrt{p^2 + m_{\text{th}}^2}} + \mathcal{O} (\Gamma_p).
\end{split}
\end{equation}
Since $m_{\text{th}} \gg \Gamma_p$, we neglect the second term in Eq.~(\ref{eq:omega}) and obtain 
\begin{equation}
\begin{split}
\langle  \partial_{\mu} \chi_1 (x) \partial^{\mu} \chi_1 (x) \rangle
&\simeq \frac{m_{\text{th}}^2}{2 \pi^2}\int_0^{\infty} \mathrm{d} p~\frac{p^2}{\sqrt{p^2 + m_{\text{th}}^2}}   \left( \frac{1}{2} + n_B \left( \sqrt{p^2 + m_{\text{th}}^2} \right) \right) \\
&= m_{\text{th}}^2 \left( \langle \chi_1^2(x) \rangle_{\text{vac}} + \langle \chi_1^2(x) \rangle_{\text{T}} \right).
\end{split}
\end{equation}
Here, $\langle \chi_1^2(x) \rangle_{\text{vac}}$ and $\langle \chi_1^2(x) \rangle_{\text{T}}$ are given by
\begin{equation}
\begin{split}
&\langle \chi_1^2(x) \rangle_{\text{vac}} = \frac{1}{4 \pi^2}\int_0^{\infty} \mathrm{d} p~\frac{p^2}{\sqrt{p^2 + m_{\text{th}}^2}}, \\
&\langle \chi_1^2(x) \rangle_{\text{T}} = \frac{1}{2 \pi^2}\int_0^{\infty} \mathrm{d} p~\frac{p^2}{\sqrt{p^2 + m_{\text{th}}^2}}~n_B \left( \sqrt{p^2 + m_{\text{th}}^2} \right)
= \frac{T^2}{2 \pi^2} J \left( \beta  m_{\text{th}} \right),
\end{split}
\end{equation}
and $J (\alpha)$ is defined by
\begin{equation}
\begin{split}
J (\alpha) \equiv \int_{\alpha}^{\infty} \mathrm{d} x~\frac{\sqrt{x^2 - \alpha^2}}{\mathrm{e}^x - 1}.
\end{split}
\end{equation}
Therefore, neglecting the vacuum contribution $\langle \chi_1^2(x) \rangle_{\text{vac}}$ 
\footnote{Since supergravity framework has a cutoff scale $M_{\text{P}}$, we can regulate the divergent vacuum contribution $\langle \chi_1^2 (x) \rangle_{\text{vac}}$. We find that the temperature dependent part of $\langle \chi_1^2 (x) \rangle_{\text{vac}}$ has a form like $\frac{m^2_{\text{th}}}{8 \pi^2} \mathrm{ln} \frac{m_{\text{th}}}{2 M_{\text{P}}}$ and is much smaller than $\langle \chi_1^2 (x) \rangle_T$ for a sufficiently small coupling $g$.}, we obtain
\begin{equation}\label{eq;kin-realres}
\begin{split}
\langle  \partial_{\mu} \chi_1 (x) \partial^{\mu} \chi_1 (x) \rangle
&= \frac{m_{\text{th}}^2 T^2}{2 \pi^2} J \left( \beta m_{\text{th}} \right) \\
&\simeq \kappa \frac{g^2 T^4}{12}.
\end{split}
\end{equation}
Here, since the thermal mass of $\chi_1$ depends on the form of interaction with particles in the thermal bath, 
we have introduced a model-dependent parameter $\kappa \lesssim \mathcal{O} (1)$ and used thermal mass\footnote{
For example, if the interaction term is given by $\mathcal{L}_{\text{int}} = - \frac{g}{\sqrt{2}} \chi_1 \bar \psi \psi - \frac{g^2}{2} \chi_1^2 (\tilde \psi_1^* \tilde \psi_1 + \tilde \psi_2^* \tilde \psi_2)$,
the parameter is $\kappa = 1/4$ in 1-loop Hard-Thermal-Loop (HTL) approximation~\cite{Bellac}.
Here, $\psi, \tilde \psi_i~(i=1,2)$ are massless Dirac Fermion and complex scalar field, respectively, and both in the thermal bath.
In finite-temperature with HTL approximation, a fermionic loop has not only the same factor $(-1)$ as in the zero temperature system
but also has another factor $(-1)$ arising from the anti-periodicity of fermionic field~\cite{Bellac}.
Thus, the bosonic and fermionic contributions to the thermal mass of $\chi_1$ do not cancel out each other. 
} 
of the form $m_{\text{th}}^2 = \kappa g^2 T^2$.
We have also used an approximation $J \left( \beta m_{\text{th}} \right) = J \left( \kappa g \right) \sim J (0) = \pi^2 /6$ in the last line.

Then, we can conclude that the statement ``$\langle  \partial_{\mu} \chi_1 (x) \partial^{\mu} \chi_1 (x) \rangle \sim g^2 T^4$'' for the real scalar field $\chi_1$ is verified 
in the case where the width is much smaller than the quasiparticle energy.
It is easy to extent Eq.~(\ref{eq;kin-realres}) to the kinetic term for the complex scalar field $\chi$:
\begin{equation}
\begin{split}
\langle  \partial_{\mu} \chi^* (x) \partial^{\mu} \chi (x) \rangle 
&= \langle  \partial_{\mu} \chi_1 (x) \partial^{\mu} \chi_1 (x) \rangle  \\
&= \frac{m_{\text{th}}^2 T^2}{2 \pi^2} J \left( \beta m_{\text{th}} \right).
\end{split}
\end{equation}
Then, the effective mass-squared for the scalar field $\phi$ contributed from the kinetic term of $\chi$ is given by
\begin{equation}\label{eq:result}
\begin{split}
\tilde m_{\phi}^2 |_{\text{kin.}}
&=  - \frac{c}{M_{\text{P}}^2} \frac{m_{\text{th}}^2 T^2}{2 \pi^2} J \left( \beta m_{\text{th}} \right) \\
&\simeq  - \frac{15 c \kappa}{2 \pi^2 g_*} g^2 H^2,
\end{split}
\end{equation}
where the relation $3 M_{\text{P}}^2 H^2 = \frac{\pi^2 g_*}{30} T^4$ in RD era is used.
Here, $g_*$ is the effective number of the relativistic degrees of freedom in the thermal bath.
This is the result which answer the question we raise in Introduction, namely, 
the thermal plasma in the early universe provide a source for the Hubble-induced mass-squared $\simeq g^2H^2/g_*$ under the K$\ddot{\rm{a}}$hler potential Eq.~(\ref{eq:kahler}).
We note that if there are $N$ complex scalar fields like $\chi$ in the thermal bath,
the Hubble-induced mass-squared Eq.~(\ref{eq:result}) would be enhanced by a factor $N$.

So far, we have considered the kinetic term of the scalar field $\chi$ only.
In supersymmetry framework, however, there is also fermionic counterpart $\tilde \chi$.
It is expected that the kinetic term of $\tilde \chi$ also contributes to the effective mass-squared $\tilde m_{\phi}^2$.
Using KB equations, we could evaluate the contribution from fermion kinetic term as well as the bosonic one.
The result  will be reported elsewhere~\cite{preparation}.

Finally, we comment on the effects of the superpotential.
In addition to the kinetic term, the K$\ddot{\rm{a}}$hler potential Eq.~(\ref{eq:kahler}) is also coupled to scalar and fermionic fields in the superpotential.
So, superpotential may provide another significant source to the effective mass-squared $\tilde m_{\phi}^2$.
Such a contribution would give the same order of Eq.~(\ref{eq:result}).

\section{Conclusion}
\label{sec:conc}
We have analyzed the effect of the non-minimal K$\ddot{\rm{a}}$hler potential Eq.~(\ref{eq:kahler}) in RD era.
In order to base on a reliable formalism, we use KB equations Eq.~(\ref{eq:12KB}) to express the expectation value $\langle  \partial_{\mu} \chi^* \partial^{\mu} \chi \rangle$
despite the fact that we are interested in equilibrium system.
The result we have obtained under reasonable assumptions is given in Eq.~(\ref{eq:result}),
which makes sure the existence of order $H^2$ contribution to $\tilde m_{\phi}^2$ in RD era.
Such a Hubble-induced mass-squared in RD era may affects some cosmological scenarios.
A complete analysis which also includes other possible sources  will be provided in Ref.~\cite{preparation}.
Although we have assumed $\chi$ is in thermal equilibrium in this study,
the time evolution of the effective mass-squared $\tilde m_{\phi}^2$ could be investigated by using KB equations to a non-equilibrium situation
under a reasonable definition of the effective mass.

\section*{Acknowledgments}
This work is supported by Grant-in-Aid for Scientific research from
the Ministry of Education, Science, Sports, and Culture (MEXT), Japan,
No.\ 14102004 (M.K.), No.\ 21111006 (M.K.) and also 
by World Premier International Research Center
Initiative (WPI Initiative), MEXT, Japan. 

{}

\end{document}